\documentclass{optica-article}

\journal{opticajournal} 

\articletype{Research Article}
\usepackage{caption}
\usepackage{subcaption}
\usepackage{lineno}

\begin{document}

\title{Geometry Driven Spin Polarization Effects in Waveguiding Plasmonic Crystal}

\author{Suman Mandal,\authormark{1} Apuroop Vaidyam,\authormark{1} Nishkarsh Kumar,\authormark{2} Sujit Rajak, \authormark{1}, Shyamal Guchhait, \authormark{3,*}  and Nirmalya Ghosh\authormark{1,*}}

\address{\authormark{1}Department of Physical Sciences, Indian Institute of Science Education and Research Kolkata, Mohanpur 741246, India\\
\authormark{2}Department of Physics, Indian Institute of Technology Kanpur, Uttar Pradesh 208016, India\\
\authormark{3}Department of Bioengineering, University of Washington, Seattle, Washington 98195, USA}

\email{\authormark{*}nghosh@iiserkol.ac.in, shyamal@uw.edu} 


\begin{abstract*} 
We investigated the modulation in the polarization-dependent optical behaviour of the waveguiding plasmonic crystal by varying the illumination and detection geometry. We employed the finite element method-based COMSOL simulation and Jones-Mueller formalisms to probe the variations in optical parameters by systematically varying the angle of incidence and azimuthal angle of the incident plane wave source, thereby quantifying the variations in the polarization anisotropy parameters. The enhancement of optical properties at various resonances and the tunability of the hybridized modes are discussed. Importantly, our study reveals that despite the system being perfectly inversion-symmetric and achiral, it exhibits circular polarization anisotropy effects for non-zero finite azimuthal angles, manifested in the off-block-diagonal elements of the constructed Mueller matrices. The origin of this intriguing circular anisotropy effect is unravelled using the sequential linear birefringence and linear diattenuation effects arising from geometrical polarization transformation. Its implications in the spin-orbit interaction of light in the plasmonic crystal system are discussed.

\end{abstract*}

\section{Introduction}
The ability to control light at subwavelength scales has rapidly advanced nanophotonics, with metasurfaces and plasmonic waveguide systems emerging as powerful means to manipulate the amplitude, phase, and polarization of light. It is also well known that light carries linear momentum\cite{jackson2012classical} along with two kinds of angular momenta, orbital (OAM)\cite{allen1992orbital} and spin (SAM)\cite{beth1936mechanical}. While these spatial and polarization degrees of freedom are independent in the case of a planar beam, they become coupled when considering field confinement in a narrow space, resulting in Spin-Orbit Interaction (SOI) effects\cite{bliokh2015spin,soumyanarayanan2016emergent,gupta2015wave}. SOI has been studied in various systems, such as the tight focusing of light beams, propagation through inhomogeneous anisotropic media, and different nano-optical systems, including metasurfaces and wave-guiding plasmonic crystals\cite{liang2024polariton,wolf1959electromagnetic,nayak2023spin,gupta2015wave}. In most cases, however, the SOI effects are weak. SOI has potential applications in spin-momentum locking, spin-controlled directional waveguiding, transverse SAM\cite{rodriguez2013near, petersen2014chiral, pan2016strong,aiello2015transverse} etc. Consequently, there is a growing need to enhance these effects due to the demand for spin-controlled devices.

 By exploiting resonant interactions and strong field confinement in nanostructures, hybridized systems such as waveguiding plasmonic crystals (WPCs) provide an effective platform for enhancing spin–orbit interaction. WPCs have the extraordinary capability of coupling the plasmon resonance and waveguide resonance in the near-field, leading to strong confinement of electric field, and  the near-field coupling and hybridization of modes is manifested in the far-field  in intriguing effects such as optical Fano resonances\cite{fano1961effects,ray2017polarization}, electromagnetically induced transparency, and electromagnetically induced absorption\cite{lu2012plasmonic,wan2015electromagnetically,taubert2012classical}, SOI etc. Additionally, WPCs have also been utilized in ultra-sensitive sensors for chemical and biosensing\cite{ray2022fano,guchhait2024metal,barshilia2024waveguide}, on-chip spectral filters\cite{ebadi2024design}, and to enhance nonlinear effects\cite{zhang2025efficient}, among numerous other applications.\par                

In a recent study\cite{nayak2023spin} on waveguiding plasmonic crystal, we have observed a very intriguing manifestation of enhanced SOI effect. It was shown that tight focusing onto a  WPC structure generates a spatially varying geometrical phase, and its interaction with the anisotropic WPC system leads to spin (circular polarization)-dependent polarization effects in the form of both forward and inverse spin Hall effects of light. While these results were qualitatively explained, a comprehensive quantitative understanding of these unusual SOI effects has yet to be established. In order to address this, here we have investigated the origin of the spin-dependent scattering effects in the inversion symmetric and achiral WPC system and extracted the corresponding geometrical polarization anisotropy parameters through numerical simulations of the interaction of polarized light with the WPC system.\par For this purpose, the crystal structure is divided into smaller subdomains, and Maxwell’s equations are solved using the COMSOL solver based on finite element method \cite{COMSOL_WaveOptics}, with appropriate electromagnetic boundary conditions imposed on each subdomain. The system is simulated under plane-wave illumination across a broad wavelength region, while systematically varying the incidence angle and azimuthal angle for different input polarization states. We have constructed the Mueller-Jones matrices for the interaction of polarized light with the WPC system and studied the resulting SOI phenomena reflected in the off-diagonal   of the computed Mueller matrices. We have observed that for input wave vector having a non-zero azimuthal angle, spin (circular polarization)-dependent polarization effects appear counterintuitively in the circular anisotropy-descriptor Mueller matrix elements, despite the WPC being completely inversion-symmetric and having no intrinsic structural chirality.  We unravel the geometric origin of this circular anisotropy effect emerging due to sequential polarization transformations in the plasmonic and waveguiding subsystems, and discuss its connection with SOI of light.\par 

\section{Theory \& Simulation}
\begin{figure}
    \centering
\includegraphics[width=1 \linewidth]{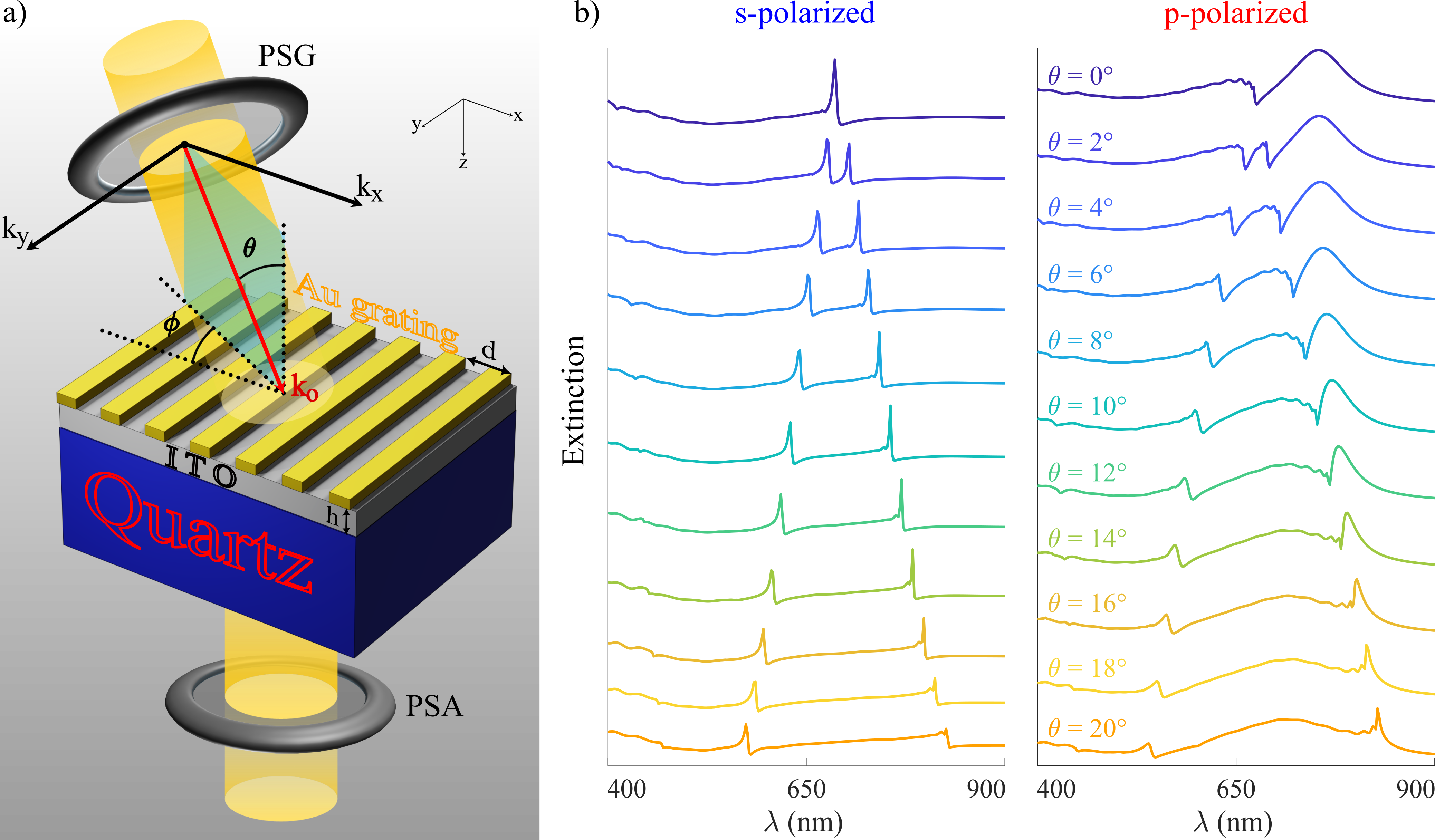}
    \caption{
\textbf{Schematic representation of the simulation geometry for scattering of light from waveguiding plasmonic crystal.} 
(a) Schematic of the simulation geometry. The incident wave vector \(\vec{k}_{0}\), shown in red, is defined by the incidence angle \(\theta_{i}\) and azimuthal angle \(\phi\). PSG and PSA denote the polarization state generator and analyzer, respectively. The gold grating (period \(d_{x} = 450\,\text{nm}\)) is placed on top of an indium tin oxide (ITO) waveguide layer of height \(h = 140\,\text{nm}\), supported by a quartz substrate. 
(b) Computed extinction spectra for s (senkrecht) and p (plane) polarized light in the wavelength range \(400\text{–}900\,\text{nm}\). From top to bottom, the incidence angle \(\theta\) is varied from \(0^{\circ}\) to \(20^{\circ}\) in increments of \(2^{\circ}\), while the azimuthal angle is fixed at \(\phi = 0^{\circ}\). The spectra are shifted vertically to facilitate comparison.
}
  \label{Fig1}
\end{figure}

We have investigated the polarization effects in a WPC composed of a periodic gold Grating (gold nano-wire array) that is exposed to air and deposited on top of an Indium tin oxide (ITO) layer which in turn is placed on top of a quartz substrate. The gold grating has a periodicity of 450nm, the nanowires' height and width being 20nm and 100nm respectively in our system. This metal-dielectric interface supports plasmon excitations upon satisfying appropriate momentum matching conditions\cite{maier2007plasmonics}.
The ITO layer in the system acts as the waveguide and has a thickness of 140nm. 
These dimensions are chosen specifically to excite plasmons in the desired spectral region with a reasonable lineshape, ensuring that both plasmonic and waveguide modes are manifested with a strong overlap between them. The thickness of the waveguiding layer is chosen to be neither too small nor too large so that it is capable of supporting a waveguide mode, but does not support the higher-order waveguide modes, which would needlessly complicate the study as well as the analysis. To study the polarization response, we generate two orthogonal polarization states (s and p polarizations can be experimentally generated with a PSG), that are made incident on the WPC with the momentum $\vec{k}$ which is considered to make an angle of $\theta_{i}$ with the normal and an azimuthal angle $\phi_{i}$ with direction of grating periodicity. The transmitted light is collected past the quartz substrate, and the polarization states are analyzed (can be done experimentally by placing a PSA here).
The system, incidence, and collection geometry are presented in figure \ref{Fig1}a.\par
This system is mimicked using the COMSOL Wave Optics module, where we have constructed a unit cell, that is, a single gold nano-wire (shown in Supplement section S1, Fig.S1). The simulation is processed using the Finite Element Method (FEM) that discretizes this unit cell by construction of a mesh of the object. For efficient computing, a finer mesh is used in the region surrounding the gold nano-wire and the ITO in comparison to the enclosing media. Given a set of initial and boundary conditions, the Maxwell's equations are solved numerically across the mesh boundaries.  In our context, to simulate the nano-wire array, it suffices to just consider the singular unit cell and apply Floquet periodicity across the boundaries.
The grating periodicity is taken along the $\hat{x}$ direction, and the length of the grating is taken along the $\hat{y}$ direction. And so, the structure is homogeneous along the $\hat{y}$ direction. The system is illuminated from the top with a plane wave having wavelength in the range of 400 nm to 900 nm. Incident light with electric field directed along the length of the grating is denoted as Transverse Electric (TE) polarization while incident light with electric field direction parallel to the direction of periodicity is denoted as Transverse Magnetic (TM) polarized light according to the convention used by A. Christ et. al \cite{PhysRevB.70.125113}. However, as our study includes cases with varying azimuthal angle $\phi_{i}$, we will use p-polarized and s-polarized to denote electric field directions parallel and perpendicular to the plane of incidence respectively. As such, p-polarized light would be TM-like for $\phi_{i}=0^o$ while it would be TE-like for $\phi_{i}=90^o$  and vice-versa for s-polarized.
The structure is illuminated from the top with a plane wave having wavelength in the range 400nm - 900nm. The structure is homogeneous along the $\hat{y}$-direction and it is illuminated from the top, therefore, energy flows along the $\hat{z}$-direction. So, for a given incidence geometry $\theta_{i}$ and $\phi_{i}$, the electric field of the incident light undergoes a transformation given by
\begin{equation}
\vec{E}_{global} = R_{z}(-\phi) \cdot R_y(\theta) \cdot \vec{E}_{in}
\end{equation}
Where $\vec{E}_{in}$ is the electric field vector in the local coordinate system of the incident light, which coincides with the global coordinate system at normal incidence, and $R_y(\theta_{i}),R_z(-\phi)$ are the general rotation matrices. For the numerical simulation, we have taken the quartz substrate thickness to be 2.7$\mu$m. As the output port is at the bottom of the substrate, this thickness ($\sim 3\lambda_{max}$) is sufficient to reduce near-field effects while being small enough to not generate any computing artefacts of higher modes supported by quartz. The dielectric air column placed above the ITO, enclosing the grating is of height 1$\mu$m. To ensure unidirectional energy flow, we have considered two perfectly matched layers (PMLs) of thickness 200nm each, one atop the air layer and another beneath the substrate. Optical properties of the grating material Gold, are defined using COMSOL's built-in material library following the model of Rakic et. al \cite{rakic1998optical}. In a similar manner, material properties of the ITO are defined using the model of Moerland and Hoogenboom \cite{Moerland:16}, while the refractive index of glass is user-defined, where we have specified the real part of the refractive index of glass to be 1.5 and the imaginary part of the refractive index to be 0.  \par

Projecting two orthogonal polarization states (s and p) of light and recording the phase and amplitude of the corresponding scattered electric field at the bottom of the quartz substrate (described in Supplementary Section S1), we constructed the Jones matrices corresponding to all wavelengths in the range 400 nm-900 nm using the relation \cite{gupta2015wave} given by
\begin{align}
\begin{bmatrix}
        E_{x,out}\\
        E_{y,out}\\
\end{bmatrix}
    =
\begin{bmatrix}
J_{11} & J_{12} \\
J_{21} & J_{22}
\end{bmatrix}
\begin{bmatrix}
        E_{x,in}\\
        E_{y,in}
\end{bmatrix}
\end{align}
Since the system of our study is non-depolarizing, we constructed the Mueller matrix and its wavelength variations from the Jones matrix using the relation shown below. \cite{gupta2015wave}
\begin{equation}
   M=A(J \otimes J) A^{-1} 
\end{equation}
Analysing these Mueller matrices, we have subsequently quantified the parameters of our interest, such as linear diattenuation, linear retardance, their orientations, etc.\par

Further, we can express the computed Mueller matrices as the product of the diattenuator and retarder Mueller matrices following Lu-Chipman decomposition \cite{Lu:96} $M=M_{D}M_{R}$, where $M_{R}$ can be expressed as

\begin{equation}
M_{D} = T_{u}
\begin{bmatrix}
1 & \vec{D}^{\,T} \\
\vec{D} & \mathbf{m}_{D}
\end{bmatrix},
\qquad
M_{R} =
\begin{bmatrix}
1 & \vec{0}^{T} \\
\vec{0} & \textbf{m}_{LR}
\end{bmatrix}.
\end{equation}

And calculate \cite{10.1117/1.2960934} the polarization parameters linear diattenuation($d$), linear retardance($\delta$) and their axis of orientation($\theta_{D}$ and $\theta_{R}$ respectively) as
\begin{align}
d \;&= \tfrac{1}{2}\sqrt{m_{12}^{2}+m_{13}^{2}}
\qquad
& \theta_{d} \;&= \tfrac{1}{2}\tan^{-1}\!\left(\frac{m_{13}}{m_{12}}\right) \\
\delta \;&= \cos^{-1}\!\left(\frac{\operatorname{tr}(M_{R})}{2}-1\right)
\qquad
& \theta_{\delta} \;&= \tfrac{1}{2}\tan^{-1}\!\left(\frac{r_{2}}{r_{1}}\right)
\end{align}
Here, $m_{ij}$ is the Mueller-matrix element in the $i^{th}$ row and $j^{th}$ column, whereas,\newline $r_i=\frac{1}{2\sin\delta}\sum_{j,k=1}^{3}\varepsilon_{ijk}\cdot m_{LR}(j,k)$ are the components of the retardance vector \cite{Lu:96}.
\\\\Extinction coefficients corresponding to the two orthogonal input polarizations for all wavelengths have been calculated from the two-port S-parameters. As a consistency check, we computed the extinction coefficients for s and p-polarizations at $\phi_{i}=0^o$, varying the angle of incidence from $0^{\circ}$ to $20^{\circ}$ in steps of $2^{\circ}$ for wavelengths in the range 400 nm - 900 nm in steps of 2nm. The results are plotted as extinction spectra, vertically shifted to facilitate comparison, in figure \ref{Fig1}b. For the incident s-polarization (left panel of figure \ref{Fig1}b), the system only supports the waveguide modes and not the plasmon ones, while for p-polarization (right panel of figure \ref{Fig1}b), both the waveguide modes and plasmon resonance are supported, resulting in hybridized modes. Extinction plots shown in figure \ref{Fig1}(b) and figure \ref{Fig1}(c) clearly exhibit optical mode separation, which is a typical signature of a strongly coupled waveguiding-plasmonic system.
\section{Results and Discussion}
\begin{figure}
    \centering
\includegraphics[width=1 \linewidth]{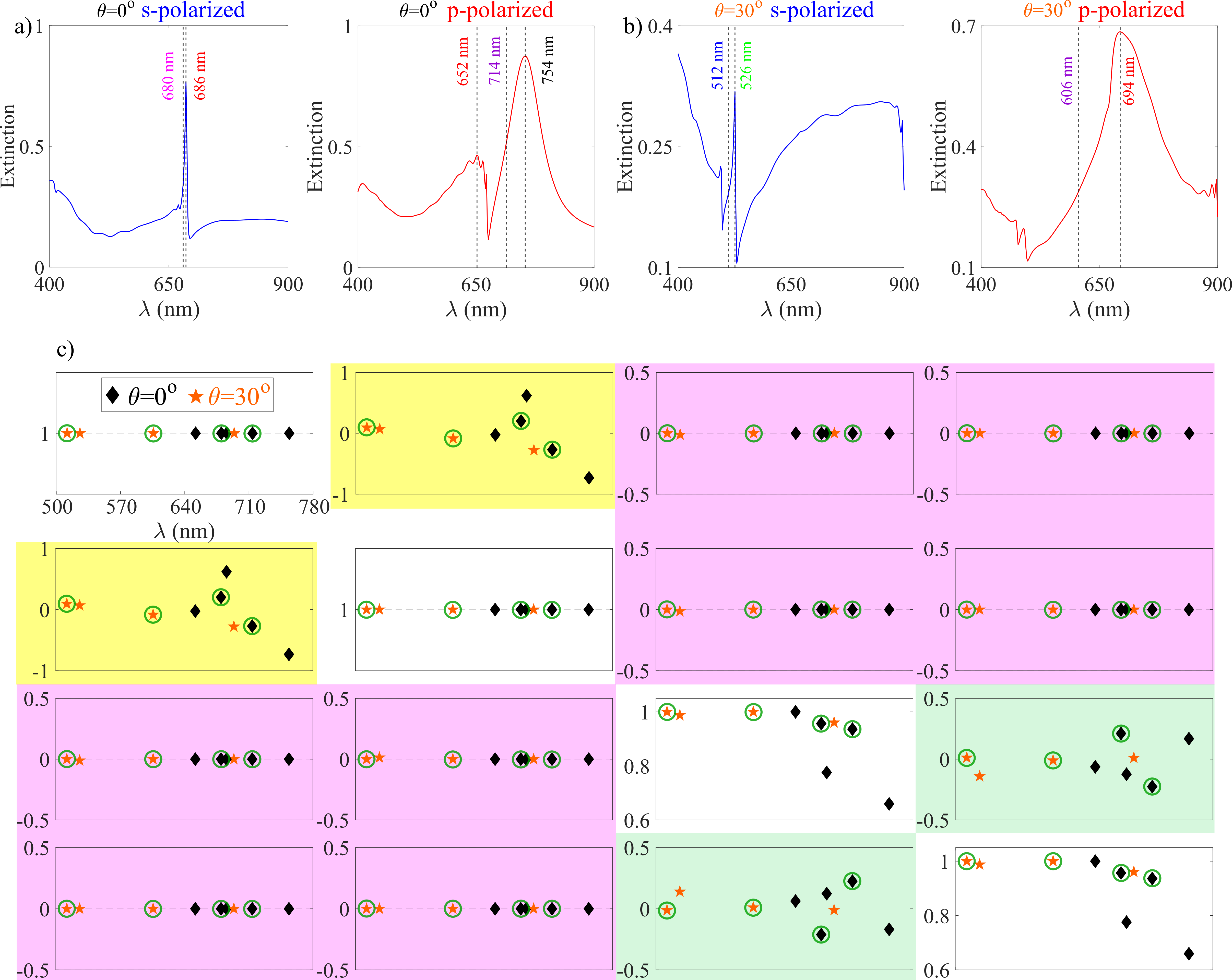}
    \caption{\textbf{Spectral polarization characteristics of the WPC system encoded in $4\times 4$ Mueller matrix for varying angle of incidence ($\theta_{i}$)}. The extinction spectra for s and p polarizations with incidence angles (a) $\theta_{i}=0^{\circ}$, $\phi_{i}=0^{\circ}$, (b) $\theta_{i}=30^{\circ}$, $\phi_{i}=0^{\circ}$. The computed Mueller matrix at the wavelengths of interest marked in (a), (b) is shown in (c): The black diamond and saffron star markers correspond to the peak wavelengths in (a) (652nm, 686nm, 754nm), (b) (526nm, 694nm) respectively, while those circled in green indicate the non-peak wavelengths in (a) (680nm, 714nm) and (b) (512nm, 606nm).}
    \label{Fig2}
\end{figure}

Our simulation is classified into two scenarios that broadly capture the polarization response of WPC when illumination by a plane wave for various incidence geometry. We vary the angle of incidence $\theta_{i}$ in one of them, while varying the azimuthal angle $\phi_{i}$ in the other.
In the first scenario, the angle of incidence is varied from $\theta_{i}=0^\circ$ to $\theta_{i}=60^\circ$ in steps of $10^\circ$, while keeping the azimuthal angle constant at $\phi_{i}=0^\circ$, and  computed the extinction spectra and Mueller matrices across the wavelength range 400nm-900nm. Of these, we have considered two cases -- $\theta_{i}=0^\circ$ and $\theta=30^\circ$. The Extinction spectra are given in figure \ref{Fig2}a and \ref{Fig2}b respectively with the wavelengths of regions of interest marked, while the Mueller matrices are shown in figure \ref{Fig2}c at the marked wavelengths, which correspond to either peaks or intermediate (circled green) regions.
\\ 
For normal incidence (figure \ref{Fig2}a), the Fano-type resonance peak of s-polarized light at $\lambda=686$nm is attributed to the TE quasiguided mode while the extinction peaks of p-polarized light at $\lambda=652$nm and $\lambda=754$nm are attributed to plasmon resonance and TM quasi-guided modes respectively. In the case of incidence geometry $\theta_{i}=30^\circ,\:\phi_{i}=0^\circ$ (figure \ref{Fig2}b), the peak at $\lambda=526$nm for s-polarized light is attributed to one of the excited TE quasiguided mode while the other has moved beyond the observed spectral range. We note a single significant peak at $\lambda=694$nm for p-polarized light attributed to the dominant hybridized waveguide-plasmon mode. One of the two peaks attributed to TM quasiguided mode has diminished while the other has moved beyond observed spectral range. This behaviour is consistent with what is seen in figure \ref{Fig1}b.

For this incidence setup ($\phi_{i}=0^{\circ}$ and varying $\theta_{i}$), we note that for all wavelengths, a clear block diagonal Mueller matrix is observed as shown in figure \ref{Fig2}c (also refer Supplementary figures S2, S3) , which implies that the system is a perfect linear diattenuating retarder as is expected from a inhomogeneous scattering  object. 
At the extinction peaks, the enhancement in linaer diattenuation is clearly seen, represented in $M_{12},M_{21}$ (shaded yellow) elements. A greater contribution of the enhancement comes from the TM hybridized mode due to the plasmon resonance. We also note a higher linear retardance in the intermediate regions(points circled green) compared to that of the peaks, as seen in elements $M_{34},M_{43}$ (shaded light green). This is because the peaks do not overlap, and at any one of the peak wavelengths, the system is strongly excited by only one of the two orthogonal polarizations. As both the polarizations do not strongly interact with the system at same $\lambda$, the phase difference and hence the linear retardance is not as large as it could have been. However, in the intermediate wavelengths between the peaks, although neither polarization excites the system as strongly as it does at the peaks, the combined interaction of the two polarizations produces a greater linear retardance.

We observe that for $\phi_{i}=0^\circ$, varying the angle of incidence $\theta_{i}$ would still preserve the block diagonal structure of the Mueller matrix. The Mueller matrix symmetry is also preserved with $M_{12}=M_{21}$, $M_{43}=-M_{34}$. Importantly, no circular effect is observed for these scenarios, indicated by $M_{14}=M_{41}=0$. The behaviour in regards to linear diattenuation and linear retardance remains consistent with the discussed case. Finally, we note that the block diagonal structure indicates that the WPC here is a purely linear diattenuating retarder system with the constituent linear retarder and linear diattenuator optic axes coinciding, and their orientation determined by the direction of the grating (along its periodicity), that is, the global $\hat{x}$ direction.

\begin{figure}
    \centering
\includegraphics[width=1 \linewidth]{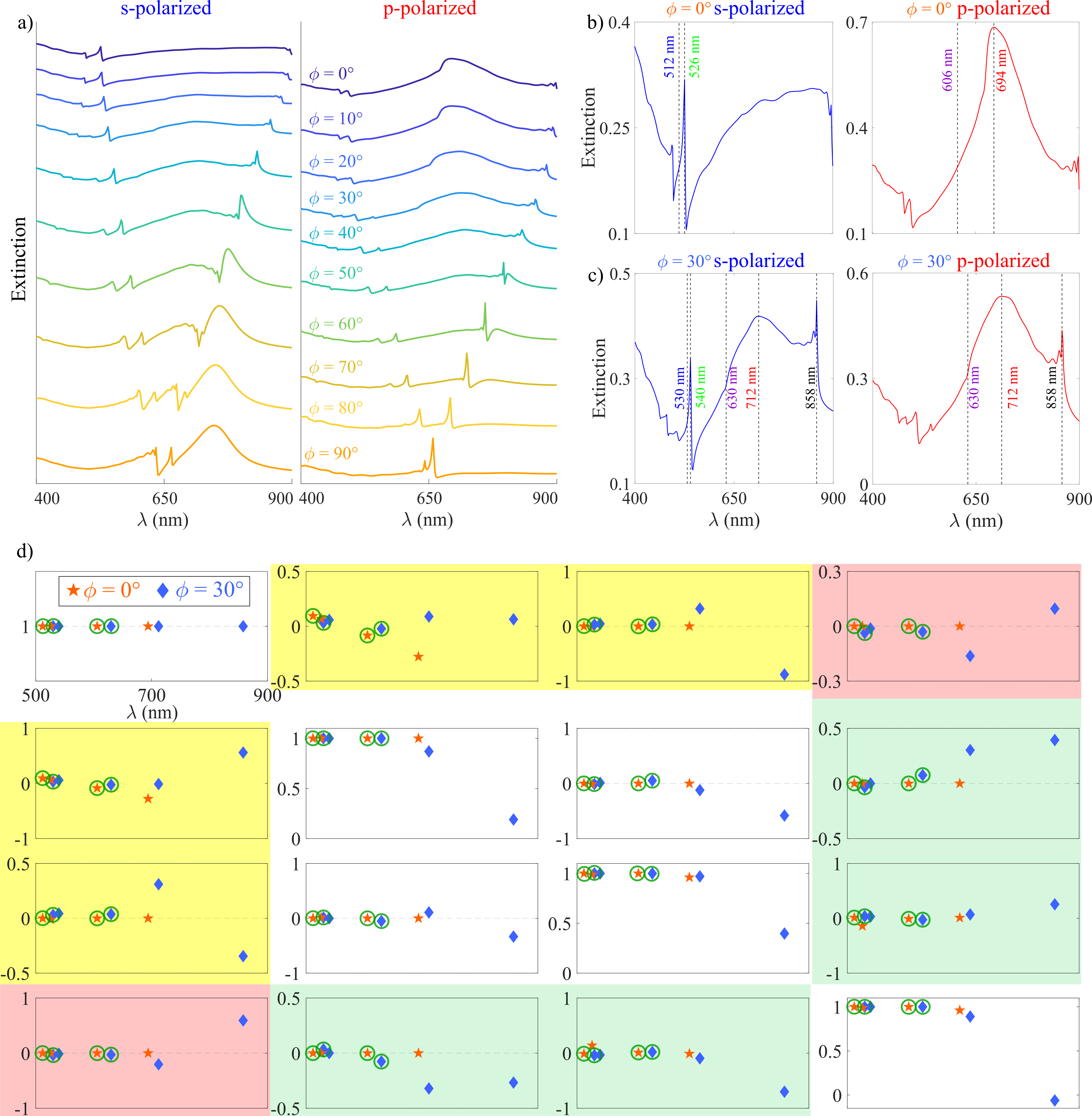}
    \caption{\textbf{Spectral polarization characteristics of the WPC system encoded in $4\times 4$ Mueller matrix for varying azimuthal angle ($\phi_{i}$)}. (a) The computed extinction spectra for s and p polarizations. From top to bottom, $\phi_{i}$ is increased from $0^{\circ}$ to $90^{\circ}$, in steps of $10^{\circ}$, while $\theta_{i}=30^\circ$ remains constant. The extinction spectra for s and p polarizations with incidence angles (b) $\theta_{i}=30^{\circ}$, $\phi_{i}=0^\circ$, (c) $\theta_{i}=30^\circ$, $\phi_{i}=30^\circ$. The computed Mueller matrix at the wavelengths of interest marked in (b), (c) is shown in (d): The saffron star and blue diamond markers correspond to the peak wavelengths (b) (526nm, 694nm), (c) (540nm, 712nm, 858nm) respectively, those circled in green indicate the non-peak wavelengths in (b) (512nm, 606nm) and (c) (530nm, 630nm).}
    \label{Fig3}
\end{figure}

In the second scenario, the angle of incidence was fixed at $\theta_{i}=30^{\circ}$ while the azimuthal angle of the incident field is systematically varied from $\phi_{i}=0^{\circ}$ to $90^{\circ}$ in steps of $10^{\circ}$ for the two orthogonal s and p input polarization states. As done in the first scenario, both the extinction and Mueller spectra are computed 
for the varying illumination geometry, which are shown in figures \ref{Fig3}a, \ref{Fig3}b and \ref{Fig3}c respectively.
The extinction spectra for the two input polarization states, show a clear azimuth dependent evolution of the WPC system, as shown in figure \ref{Fig3}a. For $\phi_{i}=0^{\circ}$, s and p polarized states correspond to TE-like and TM-like respectively, with the Fano-type resonance seen in the case of s-polarized while a broader hybrid plasmon-waveguide peak seen for p-polarized light. We observe a reversal of behavior as expected for $\phi_{i}=90^{\circ}$, with s and p polarized states corresponding to TM-like and TE-like respectively. Here, we also note the excitation of higher order modes along the gold nanowire because of the incident light momenta $k_y$, which appear as short, sharp peaks in the extinction spectra (refer Supplementary figure S5). At the intermediate azimuthal angles ($0^{\circ} < \phi_{i} < 90^{\circ}$), a mixture of TE-like and TM-like behavior is present, that is, a combination of sharp asymmetric Fano-type TE quasiguided modes, broad plasmon and TM quasiguided mode features are clearly observed in figure \ref{Fig3}a for the mentioned $\phi$(s).
We also note the clear transition of s and p polarized responses into TM-like and TE-like respectively with increasing $\phi_{i}$.

The extinction spectra of s and p polarized input states for incidence geometry $\theta_{i}=30^\circ,\:\phi_i=0^\circ$ is once again displayed in figure \ref{Fig3}b, for contrasting its behaviour with the case $\theta_{i}=30^\circ,\:\phi_i=30^\circ$, which is shown in figure \ref{Fig3}c. With the incident light momentum $\vec k$ now split along the nanowires and along the grating periodicity, the peaks shift consequently as the momenta of the excitable quasi-guided modes have changed. Here in figure \ref{Fig3}c, we clearly observe a combination of plasmon resonance, TE and TM quasiguided modes with the s-polarized case gaining broad TM quasiguided mode and plasmonic features, and the p-polarized case showing features of sharp TE quasiguided modes in contrast with the corresponding features in figure \ref{Fig3}b.

The Mueller matrices constructed for this scenario are shown in figure \ref{Fig3}d, at the wavelengths of interest marked in figure \ref{Fig3}c (full spectra shown in Supplementary figure S4). We immediately note that for $\phi_i\neq 0^\circ$ the block diagonal structure that was present in all the cases of the first scenario, has now disappeared. The previously seen matrix symmetry is also broken with $M_{12}\neq M_{21}$, $M_{43} \neq -M_{34}$ and the off diagonal elements have become non-zero. Due to this mixture of TE-like and TM-like features, we cannot distinctly attribute the optical properties to the peaks and intermediate regions. Most importantly, we note that the previously null circular diattenuation and polarizance elements $M_{14},\:M_{41}$ are now non-zero and unequal $M_{14}\neq M_{41}$. Remarkably, this indicates that the system now shows a strong circular response despite being centro-symmetric and achiral. We shift our focus to this characteristic and the dependence of the elements $M_{14},\:M_{41}$ on the azimuthal incidence angle $\phi_i$.

\begin{figure}
    \centering
\includegraphics[width=\linewidth]{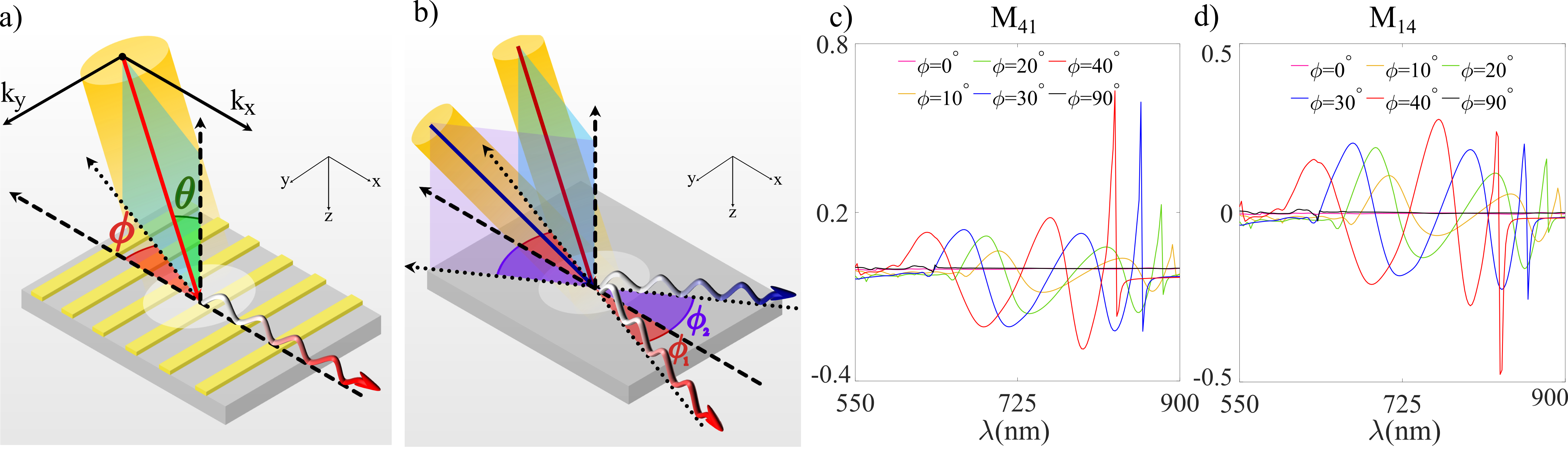}
    \caption{\textbf{Schematic illustration of the geometric origin of circular polarization (spin) dependent response from the achiral WPC system}. (a), (b) show the preferred direction of propagation of plasmon excitations and waveguide modes, respectively. The optic axis does not vary with $\phi_{i}$ in (a) for the plasmonic system but does so in (b) for the waveguide. (c), (d) show the circular polarizance and circular diattenuation spectra respectively for various incident $\phi$, with $\theta_{i}$=$30^\circ$ constant.
}
    \label{Fig4}
\end{figure}

In order to understand the presence of circular anisotropic effects in the WPC system, we propose a theoretical model where the system is composed of two linear diattenuating retarders: the gold grating and the ITO waveguide, with the Mueller matrix decomposition $M_{net}=M_{waveguide}\cdot M_{grating}$. The grating is a linear diattenuating retarder because it supports the plasmon resonance for the incident TM polarization. Similarly, the ITO waveguiding layer beneath the grating is also a linear diattenuating retarder as quasiguided modes are excited for the incident TM-TE polarizations. They are both linear because of their geometry which is centro-symmetric and achiral, that is, they are both individually, indistinguishable for the left and right circular polarizations. The combination of these two components gives a circular response when their respective optic axes do not align, which also breaks the observed block diagonal Mueller matrix structure. The axes' orientation is schematically displayed in figures \ref{Fig4}a, \ref{Fig4}b. As shown in figure \ref{Fig4}a, the orientation of the grating is fixed, there is a preferred direction of propagation of the excited plasmons and hence, the optic axis of the grating system remains unchanged under different $\phi_i$ incidence. However, there is no such preferred direction in the case of the waveguide, as shown in figure \ref{Fig4}b. This arrangement generates a circular response, due to the sequential polarization effects of the two subsystems. Specifically, $M_{waveguide},\:M_{grating}$ are both block diagonal for normal incidence which results in a net block diagonal structure with the symmetries preserved. However, when $\phi_i \neq 0^\circ$, while $M_{grating}$ remains block-diagonal, $M_{waveguide}$ does not, as it undergoes the corresponding rotational transformation (refer Supplementary Section S2).    
Hence, the resultant Mueller matrix $M_{net}$ is symmetry broken and has non-zero circular polarizance and circular diattenuation with $M_{14}\neq M_{41}$. And because of its dependence on $\phi_i$, this is a geometric effect. As further evidence,  $M_{waveguide}(\phi=90^\circ)$ is once again block diagonal and consequently, so is $M_{net}$ (refer Supplementary figure S5), supporting our mathematical formulation.

The spectral dependence of the circular polarizance ($M_{41}$) and circular diattenuation ($M_{14}$) for various incident $\phi_i$ is shown in figures \ref{Fig4}c, \ref{Fig4}d respectively. Note that these are non-zero and unequal for $\phi_i \neq 0^\circ, 90^\circ$. The shifting of the peaks with $\phi_i$ corresponds to the similar shifting of the resonant wavelengths as discussed previously for the extinction spectra. We also note that the features become increasingly pronounced with rising values of $\phi_i$, reaching a maximum before diminishing and ultimately vanishing at $\phi_{i} = 90^\circ$, as clearly illustrated in figures \ref{Fig4}c, \ref{Fig4}d. We have also analyzed the polar decomposed Mueller matrices $M_D,\: M_R$ (where $M_{net}=M_D M_R$), computing the linear diattenuation $d$, linear retardance $\delta$ and their respective orientations $\theta_d$, $\theta_\delta$. The values $d$, $\delta$ support our previous analysis and importantly, we note that the orientations $\theta_d$, $\theta_\delta$ both equal zero, that is the optic axes align along $\hat{x}$ for $\phi_i=0^\circ,\: 90^\circ$, while they do not coincide for other $\phi_i$ as seen in Table-\ref{Table-1}
.

\begin{table}[h!]
\centering

\begin{subtable}{0.45\textwidth}
    \centering
    \resizebox{\textwidth}{!}{%
    \begin{tabular}{|c c c c c c|}
    \hline
    $\theta_{\text{i}}$ & $\lambda\,(\mathrm{nm})$ & $d$ & $\delta$ & $\theta_{d}$ & $\theta_{\delta}$ \\
    \hline
    \multirow{5}{*}{$0^\circ$}
        & 652 & 0.01 & 0.42 & \multirow{5}{*}{0} & \multirow{5}{*}{0} \\
        & 680 & 0.20 & 0.22 &                    &                     \\
        & 686 & 0.62 & 0.16 &                    &                     \\
        & 714 & 0.27 & 0.24 &                    &                     \\
        & 754 & 0.73 & 0.25 &                    &                     \\
    \hline
    \multirow{4}{*}{$30^\circ$}
        & 512 &  0.09  & 0.01 & \multirow{4}{*}{0} & \multirow{4}{*}{0} \\
        & 526 &  0.07 & 0.14 &                    &                     \\
        & 606 & 0.08 & 0.01 &                    &                     \\
        & 694 & 0.28 & 0.01 &                    &                     \\
    \hline
    \end{tabular}}
    \subcaption{}
\end{subtable}
\hfill
\begin{subtable}{0.45\textwidth}
    \centering
    \resizebox{\textwidth}{!}{%
    \begin{tabular}{|c c c c c c|}
    \hline
    $\phi_{i}$ & $\lambda\,(\mathrm{nm})$ & $d$ & $\delta$ & $\theta_{d}$ & $\theta_{\delta}$ \\
    \hline
    \multirow{5}{*}{$30^\circ$}
        & 530 & 0.04 & 0.05 & 0.39 & 0.36 \\
        & 540 & 0.07 & 0.03 & 0.29 & 0  \\
        & 630 & 0.04 &  0.08 & -0.55 & 0.62 \\
        & 712 & 0.33 & 0.35 & 0.65 & -0.63    \\
        & 858 &  0.88 & 1.79 & -0.75 & -0.45 \\
    \hline
    \multirow{4}{*}{$90^\circ$}
        & 634 &  0.36  & 0.31 & \multirow{4}{*}{0} & \multirow{4}{*}{0} \\
        & 658 &  0.38 & 0.22 &                    &                     \\
        & 664 & 0.32 & 0.13 &                    &                     \\
        & 748 & 0.75 & 0.24 &                    &                     \\
        & 800 & 0.20 & 0.63 &                    &                     \\
    \hline
   
    \end{tabular}}
    \subcaption{}
\end{subtable}
 
\caption{\textbf{Quantification of anisotropy parameters (linear diattenuation $d$, linear retardance $\delta$ and their respective orientation angles $\theta_d$, $\theta_\delta$ from x-axis)  for different illumination geometries and wavelengths of interest}. (a) shows the parameters for normal and an oblique incidence at $\theta_{i}=30^{\circ}$, with constant azimuth $\phi_{i}=0^{\circ}$, for the wavelengths mentioned in figure \ref{Fig2}. (b) shows the anisotropy parameters for two azimuthal angles $\phi_{i}=30^{\circ}$, $90^{\circ}$ at constant angle of incidence, $\theta_{i}=30^{\circ}$, for the wavelengths of interests shown in figure \ref{Fig3} and supplemental material respectively. }
\label{Table-1}
\end{table}

\section{Conclusions} 
In summary, we have observed spin (circular polarization)-dependent optical effects from an inversion symmetric and achiral waveguiding plasmonic crystal (WPC) system, which arises purely from the geometrical polarization transformation in the system. For the usual normal incidence geometry, the characteristic TE and TM quasiguided resonant modes of the WPC system appear as peaks in the extinction spectra, and their behaviour under varying angle of incidence indicates a strong coupling between the waveguide and the plasmonic grating. In this configuration, the WPC acts like a pure linear diattenuating retarder, which is reflected in the characteristic block-diagonal structure of the Mueller matrix. The resonant modes exhibited strong linear diattenuation, while linear retardance was observed to be enhanced in the spectral regions between the modes. For the geometry with a non zero azimuthal angle, on the other hand, spin-dependent polarization effects are observed, manifested as appearance of non-zero off-block-diagonal elements of the Mueller matrix. Importantly, circular anisotropy emerges in the characteristic elements of the Mueller matrix as signatures of geometry-driven spin-polarization effect and unusual SOI phenomena in the inversion symmetric WPC system.  Mueller matrix analysis revealed that this unconventional SOI phenomena originates from the geometry-dependent sequential polarization transformations (sequential linear birefringence and linear diattenuation effects) generated by the plasmonic grating and the waveguiding layer.  This is shown through the azimuthal angle dependence of the orientation of anisotropy axes. Overall, this study provides  quantitative understanding and insight of an unusual SOI phenomena and their emergence in an inversion symmetric WPC system. These findings may open interesting avenues of spin-controlled nano-optical functionalities using  SOI phenomena in simple symmetric metasurfaces.  

\section{Back matter}

\begin{backmatter}
SM and SR sincerely thanks UGC and NK sincerely thanks CSIR for providing the fellowships. 

\bmsection{Acknowledgments}
SM, AV, SR and NG would like to acknowledge IISER Kolkata, NK acknowledges IIT Kanpur, and SG acknowledges the University of Washington  for the constant support. Authors sincerely thanks Dr. Jeeban Kumar Nayak and Mr. Ritwik Dhara for their insightlful suggestions.

\bmsection{Disclosures}
The authors declare that they have no conflicts of interest.

\bmsection{Data Availability Statement}
 Data underlying the results presented in this paper are not publicly available at this time but may be obtained from the authors upon reasonable request.

 \bmsection{Supplementary Information}
 See supplementary document for supporting content.
\end{backmatter}


\bibliography{sample}





\newpage
\section*{Supplementary Text}
\subsection*{S1: Simulation geometry and theoretical framework}
In this work, we have modelled the Waveguiding Plasonic Crystal (WPC) structure by creating a unit cell with the specified structure parameters, as outlined in the main text, using COMSOL's built-in CAD interface. This is followed by the implementation of a Floquet periodic boundary condition to simulate the periodic nature of the structure. As is standard in such simulations, each domain of the structure is subdivided into smaller regions, known as meshes. A user-defined mesh was employed in this simulation. In the air domain, the maximum and minimum size of mesh elements are specified as 180 nm and 18 nm, respectively. For the substrate domain, the corresponding sizes are 120 nm and 12 nm . In the waveguide domain, the maximum mesh element size is 100 nm, and the minimum element size is 10 nm. In the grating domain, which is the primary interest of our study, comparatively smaller mesh sizes are employed, with a maximum mesh element size of 20 nm and a minimum mesh element size of 5 nm. Swept mesh type is used in the perfectly matched layer (PML) domains located above the air layer and below the substrate, with 10 elements in each PML domain.  The incident electric field is input at the top of the air column and the output is averaged by a line integral that traces a circular path of radius 25nm at the bottom of the glass substrate. The schematic of the unit cell and the collection method is shown in figure S1.

\subsection*{S2: Mueller matrix modeling of the Sequential Polarization effect}
Let the Mueller matrices of the plasmonic grating and waveguiding subsystems be $M_p$ and $M_w$ respectively. These subsystems individually do not show a circular response and are pure diattenuating retarders. For a given incidence angle $\theta$, the Mueller matrix $M$ of the total system, as a function of the incident azimuthal angle $\phi$, is:   
\begin{align*}
    M(\phi)=M_w(d_w,\delta_w,\phi)*M_p(d_p, \delta_p)
\end{align*}
Where $d_{w,p}\ \&\ \delta_{w,p}$ are the linear dichroism and linear birefringence respectively for waveguiding($w$) and plasmonic($p$) subsystems. The plasmonic grating has a fixed orientation (along the $\hat{x}$ axis) and is independent of $\phi$. For a given $\theta$ and $\phi=0^\circ$ incidence, the above expression expands to:
\begin{equation} \tag{3}\label{equ:3}
M(\phi=0^\circ)=
\begin{pmatrix}
1 & d_w &0  & 0\\
d_w & 1&0& 0 \\
0 & 0 &  x_w cos\delta_w & x_w sin\delta_w\\
0 & 0 & -x_w sin\delta_w & x_w cos\delta_w
\end{pmatrix}
\begin{pmatrix}
1 & d_p &0  & 0\\
d_p & 1&0& 0 \\
0 & 0 &  x_p cos\delta_p & x_p sin\delta_p\\
0 & 0 & -x_p sin\delta_p & x_p cos\delta_p
\end{pmatrix}
\end{equation}
Where  $x_\mu = \sqrt{1-d_\mu^2}$. Here, $M_w$ and $M_p$ are block diagonal because both their orientation axes align along the chosen $\hat{x}$ axis. Hence, the resulting $M$ is also block diagonal, exhibiting features characteristic of a pure diattenuating retarder. $M_w$ however, is dependent on the azimuth $\phi$ as the waveguide has no preferred orientation axis and it undergoes the usual rotational transformation:
\begin{equation} \tag{2}\label{equ:2}
M_w(d_w,\delta_w,\phi) =
\begin{bmatrix}
1 & d_w\cos(2\phi) & d_w\sin(2\phi) & 0 \\[10pt]

d_w\cos(2\phi) &
\begin{aligned}[c]
&\cos^2(2\phi) \\[-2pt]
&+\, x_w\cos(\delta_w)\sin^2(2\phi)
\end{aligned} &
\begin{aligned}[c]
&\frac{\sin(4\phi)}{2}(1-x_w\cos(\delta_w)) \\[-2pt]
\end{aligned} &
-\,x_w\sin(\delta_w)\sin(2\phi) \\[14pt]

d_w\sin(2\phi) &
\begin{aligned}[c]
&\frac{\sin(4\phi)}{2}(1-x_w\cos(\delta_w)) \\[-2pt]
\end{aligned} &
\begin{aligned}[c]
&\sin^2(2\phi) \\[-2pt]
&+\, x_w\cos(\delta_w)\cos^2(2\phi)
\end{aligned} &
x_w\sin(\delta_w)\cos(2\phi) \\[14pt]

0 &
x_w\sin(\delta_w)\sin(2\phi) &
-\,x_w\sin(\delta_w)\cos(2\phi) &
x_w\cos(\delta_w)
\end{bmatrix}  
\end{equation}
Clearly, the resultant Mueller matrix $M=M_w*M_p$ is no longer block diagonal and importantly, its circular diattenuation and polarizance are no longer zero:
\begin{align*}
M_{14} &= x_pd_w\sin(2\phi)\sin(\delta_p)\\
M_{41} &=x_wd_p\sin(2\phi)\sin(\delta_w)
\end{align*}
Note that $M_{14}\neq M_{41}$, which supports our simulated results and discussion in the main text.

\section*{Supplementary Figures}
\renewcommand{\figurename}{\textbf{Fig.}}
\renewcommand{\thefigure}{\textbf{S\arabic{figure}}}
\setcounter{figure}{0}

\begin{figure}[]
    \centering
    \includegraphics[width=\linewidth]{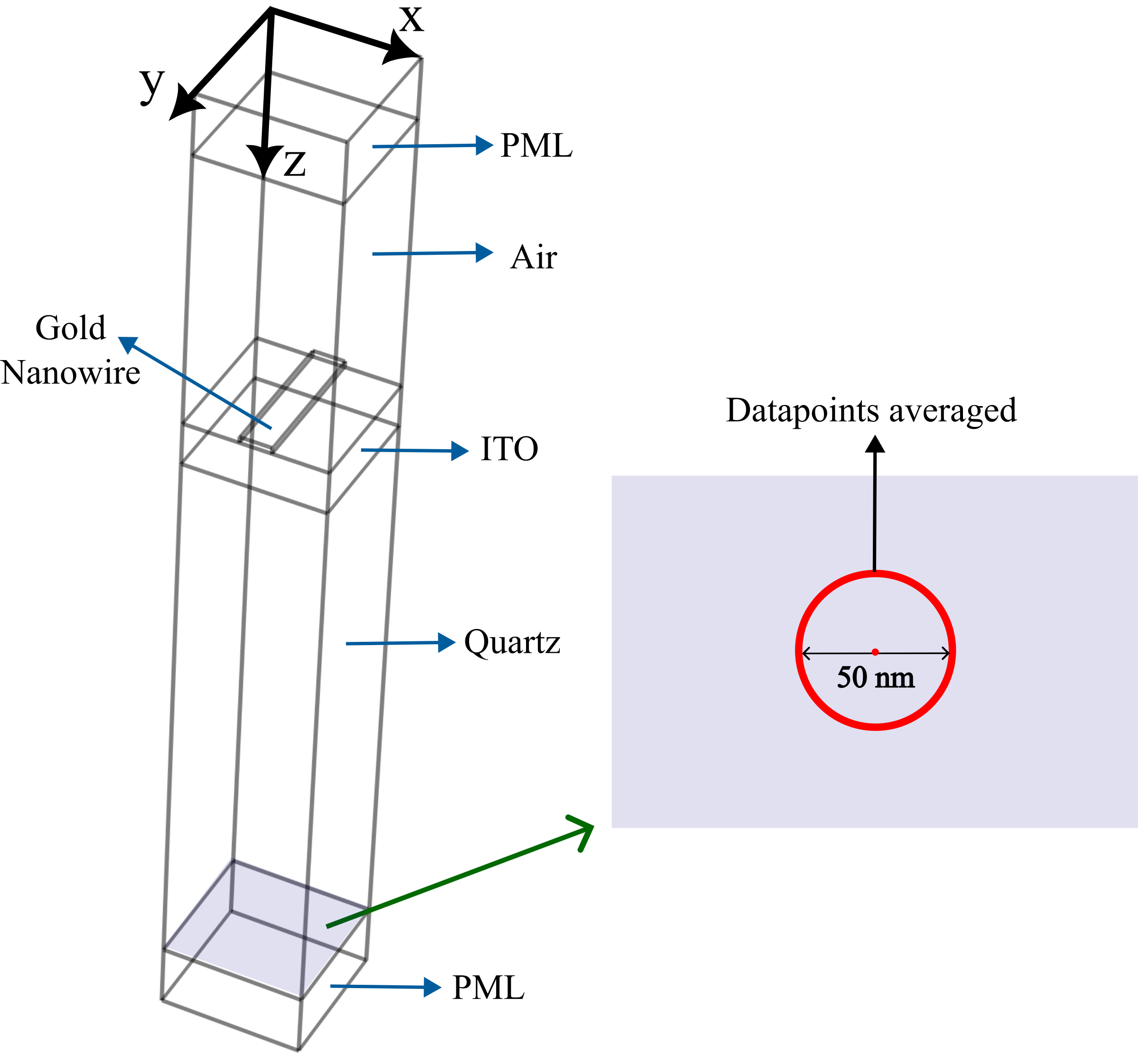}
    \caption{Schematic of a unit cell of the waveguiding plasmonic crystal with individual regions labeled is shown to the left. To its right is the averaging method shown in detail, for obtaining the output electric field.}
    \label{fig:comsol}
\end{figure}

\begin{figure}[]
  \centering
   \includegraphics[width=\linewidth]{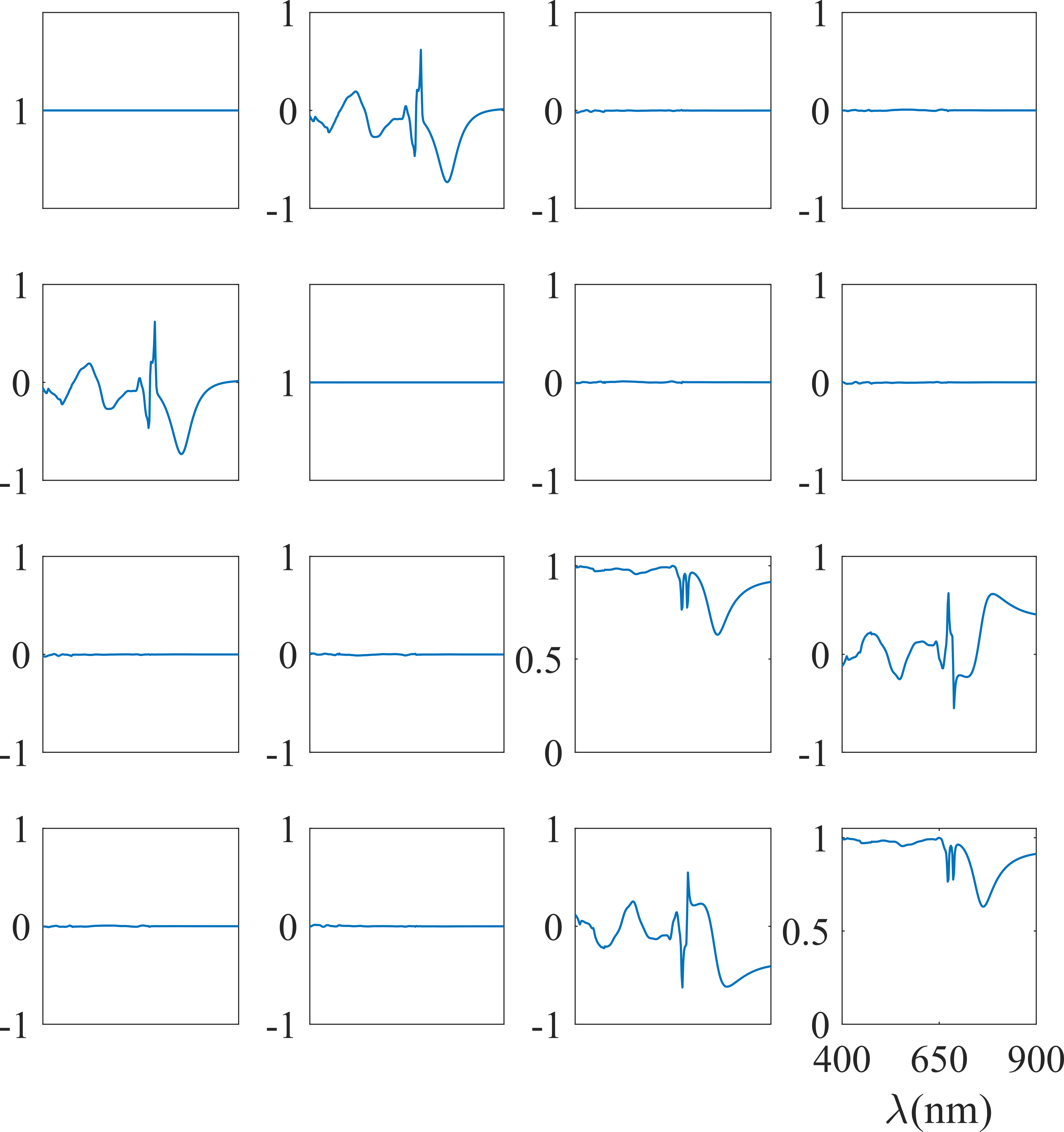}
  \caption{Spectral variation of the  Mueller matrix elements for the entire spectral wavelength range from 400 nm to 900 nm for incident angle $\theta_{i}=0^{\circ}$ and azimuthal angle $\phi_{i}=0^{\circ}$.}
  \label{Figure-1}
\end{figure}


\begin{figure}[]
  \centering
  \includegraphics[width=\linewidth]{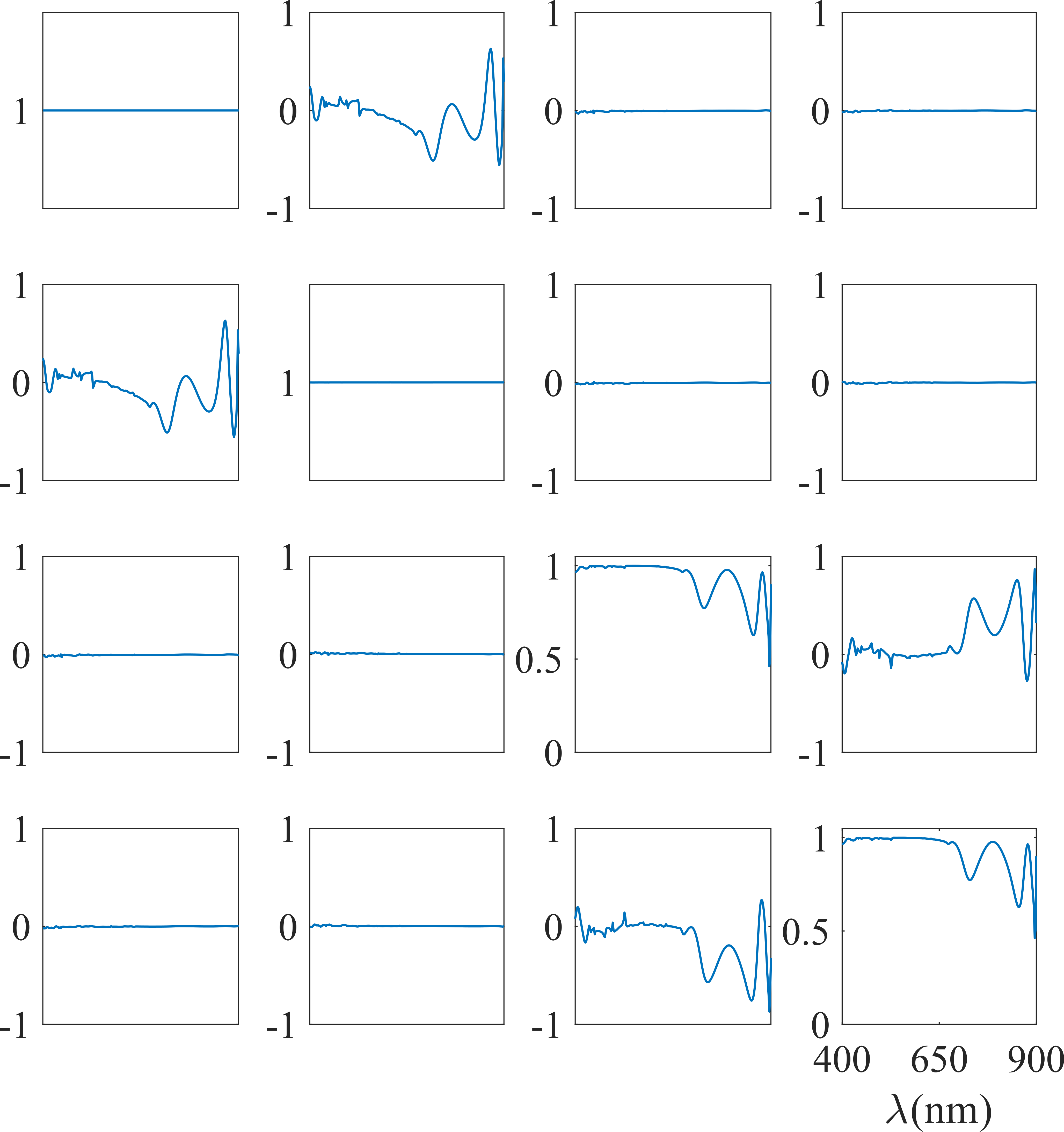}
  \caption{Spectral variation of the  Mueller matrix elements for the entire spectral wavelength range from 400 nm to 900 nm for incident angle $\theta_{i}=30^{\circ}$ and azimuthal angle $\phi_{i}=0^{\circ}$.}
  \label{Figure-2}
\end{figure}


\begin{figure}[]
  \centering
   \includegraphics[width=\linewidth]{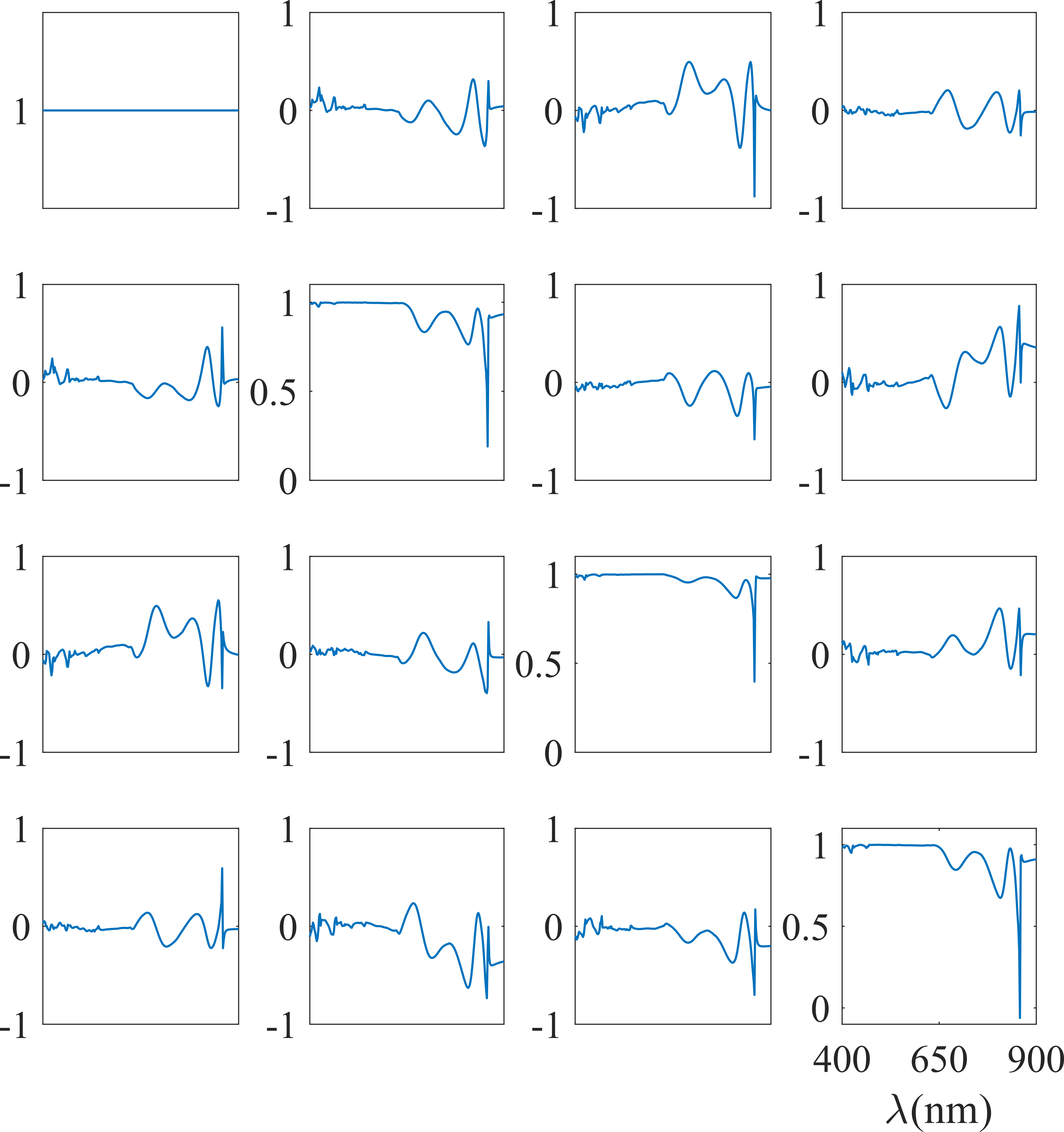}
  \caption{Full Mueller spectra for the entire spectral wavelength range from 400 nm to 900 nm for incident angle $\theta_{i}=30^{\circ}$ and azimuthal angle $\phi_{i}=30^{\circ}$.}
  \label{Figure-3}
\end{figure}


\begin{figure}
    \centering
     \includegraphics[width=\linewidth]{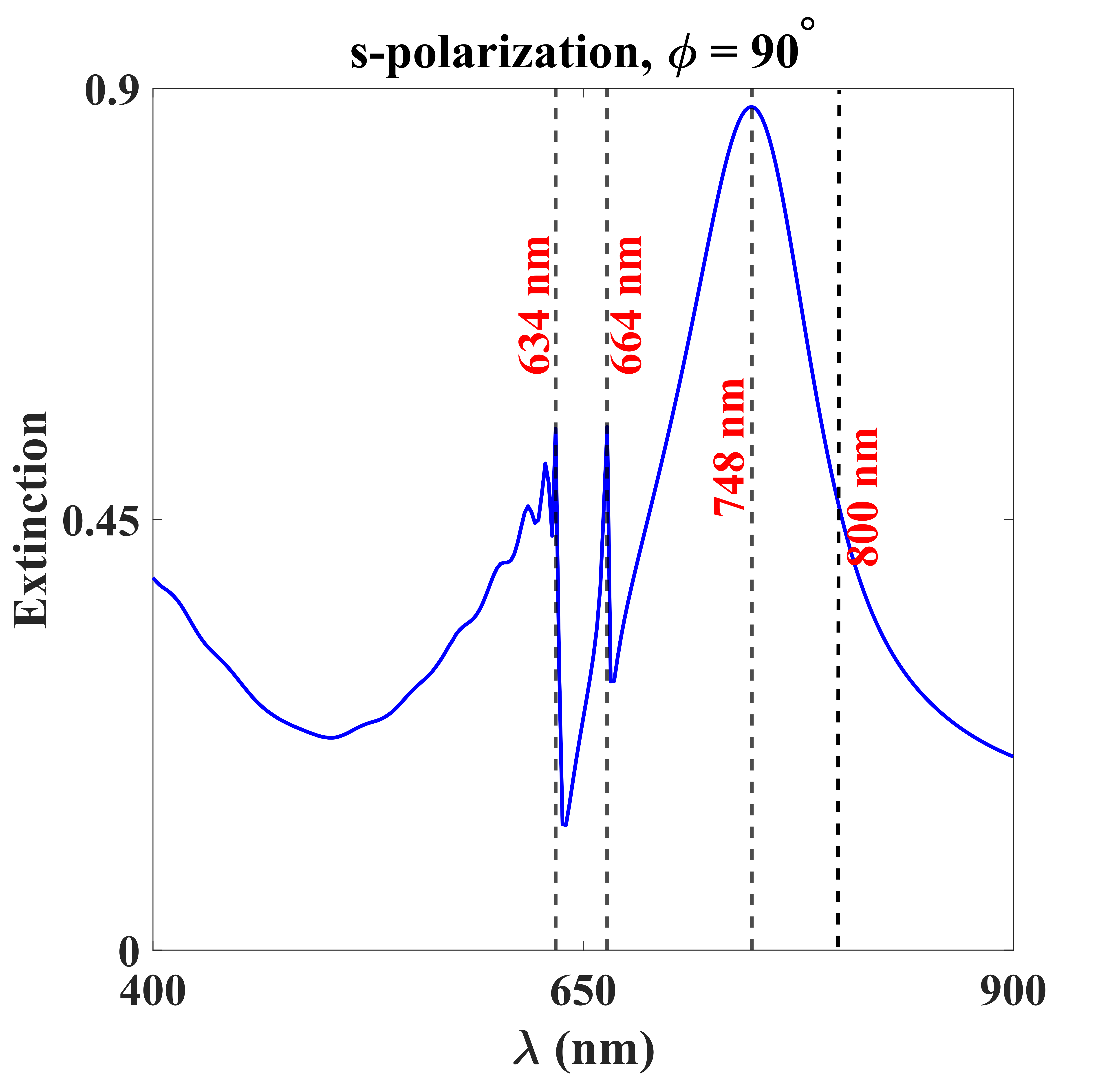}
    \caption{Extinction spectra for the two orthogonal incident polarizations, s-polarization (top-left) and p-polarization (top-right), along with full spectral Mueller matrix corresponding to $\theta_{i}=30^{\circ}$ and $\phi_{i}=90^{\circ}$ for $\theta=30^\circ$, $\phi=90^\circ$.}
    \label{Phi_90}
\end{figure}

\end{document}